\documentstyle[a4,11pt,multicol,rotating] {article}
\begin{document}

\setlength{\baselineskip}{16pt}

\begin{center}
{\bf\Large On the basis-set extrapolation}

\vspace{3mm}

{\bf\large Suresh Chandra$^{1,2*}$ \& Mohit K. Sharma$^3$}
\end{center}

\vspace{2mm}

\noindent
$^1$Physics Department, Lovely Professional University, Phagwara 144411, Punjab, India

\noindent
$^2$Zentrum f\"ur Astronomie und Astrophysik, Technische Universit\"at
Berlin, Hardenbergstrasse 36, D-10623 Berlin, Germany

\noindent
$^3$School of Studies in Physics, Jiwaji University, Gwalior 474011, M.P., India

\noindent
Emails: suresh492000@yahoo.co.in; mohitkumarsharma32@yahoo.in

\vspace{4mm}



{\bf A possible solution for the problem of memory-size and computer-time, is the 
 extrapolation of basis-set$^1$. This extrapolation has two exponents $\alpha$ and 
$\beta$, corresponding to the HF (reference energy) and the energy of correlations
(EC), respectively. For a given system, the exponents are taken as constant$^2$, and 
potential energy surfaces (PES) are generated. We have found that the values of 
$\alpha$ and $\beta$ are not constant, but vary from position to position in the 
system. How to deal with such situation and get very accurate PES, is discussed.
}

\vspace{4mm}

Truhlar$^1$ proposed a nice concept for extrapolation of basis-set, where the total 
energy is expressed as the sum of HF and EC. In the method, two exponents $\alpha$ and 
$\beta$, respectively, are introduced. 

The calculations based on the basis-sets are required in all branches of science and 
technology. In the Astronomy \& Astrophysics, their contribution is remarkable. It
 plays important role for analyzing spectrum, coming from the interstellar medium 
(ISM). 

For getting information about far distant cosmic objects, one depend on the radiations lying 
in the, which are generated by  the molecules, present in the object. The kinetic temperature
 in an object having molecules is few tens of Kelvin, in general. So, in most of the cases, 
scientists are concerned with the pure rotational levels.

We consider the molecule thioformaldehyde (H$_2$CS), for example, whose identification$^3$ in
 the ISM was considered as an achievement, because earlier attempts made several times 
remained unsuccessful, though they were based on very accurate laboratory studies. For 
analyzing spectrum, one considers an appropriate number of energy levels. These levels are 
connected through radiative and collisional transitions. Calculation of collisional cross 
sections is the most difficult task in the investigation. But, the scientists are interested 
to have accurate cross sections.  

The collisions are due to the most abundant molecule H$_2$. However, many scientists do not 
consider internal structure of H$_2$, and therefore they replace H$_2$ by the He atom, as 
both the H$_2$ and He have two electrons and two protons, and the interaction potential 
depends on the charges of the interacting particles. For the calculation of collisional cross
 sections, the interaction potential between H$_2$CS and He is required, for which the 
scientists are putting their best efforts. 

The interaction potential is calculated with the help of MOLPRO$^4$ by using the best method 
CCSD(T)$^5$ and the best basis-set the aug-ccpV$n$Z, where $n$ denotes the number of zeta
used and it stands for one of the D (Double-zeta), T (Triple-zeta), Q (Quadruple-zeta), 5 
(Quintuple-zeta), etc. The prefix 'aug' denotes the augmented versions of the basis set. 
With the increase of $n$, the accuracy of results increases, but the requirement of 
memory-size and computer-time increases exponentially, which is not affordable, beyond $n = 
6$ (say).

How to reach $n = \infty$ by using the values for $n = 2$ and $n = 3$ is a nice idea used
 in the basis-set extrapolation$^1$. That is, one has to calculate the HF and EC by 
using the CCSD(T) method along with the aug-ccpVDZ and aug-ccpVTZ
basis-sets, separately. With the help of the energies of these two sets, the limiting energies (corresponding to $n = \infty$) can be obtained. For a given system, the values of the
exponents $\alpha$ and $\beta$ were taken constant.

We put a question, Are the values of $\alpha$ and $\beta$ constant through out the system? 
The answer to this question can be found in the following manner. We prepare Figure 
\ref{Fig1}, where we plot $F [= (4^{-\gamma} - 2^{-\gamma})/(3^{-\gamma} - 2^{-\gamma})]$ 
versus $\gamma$. It is a nice smooth curve.

The position of He atom relative to the center-of-mass of H$_2$CS is expressed in
terms of the spherical polar coordinates ($r$, $\theta$, $\phi$). The interaction 
potential between the H$_2$CS and He is calculated for various positions ($r$, $\theta$, 
$\phi$).

We calculate HF and EC for one more basis-set, aug-ccpVQZ, {\it i.e.,} $n = 4$. Let the 
values  obtained by using the basis-sets aug-ccpVDZ, aug-ccpVTZ, aug-ccpVQZ, be denoted by 
$V_2$, $V_3$, $V_4$, respectively. We calculate the parameter $F = (V_4 - V_2)/(V_3 - V_2)$,
and from Figure \ref{Fig1}, find out $\gamma$, corresponding to this value of $F$. The
 $\gamma$ is either $\alpha$ or $\beta$, corresponding to the HF or EC, respectively. For 
better accuracy, this graphical method is replaced by the numerical interpolation. In this 
way, we find the values of $\alpha$ and $\beta$ at each position. 

We have considered 532 positions specified by ($r$, $\theta$, $\phi$), with $r$ = 2 (0.25) 6.5
 \AA, $\theta$ = 0$^\circ$ (30$^\circ$) 180$^\circ$, $\phi$ = 0$^\circ$ (30$^\circ$) 
90$^\circ$. The obtained values of $\alpha$ and $\beta$, are plotted in Figures \ref{Fig2} and
 \ref{Fig3}, respectively, as a function of position. Figures \ref{Fig2} and \ref{Fig3} show 
that neither $\alpha$ nor $\beta$ is constant. Both of them vary from position to position in 
the system. For very large $r$, the $\alpha$ and $\beta$ attain the limiting values, denoted 
by $\alpha_0$ and $\beta_0$. The values obtained are $\alpha_0 = 3.157$ and $\beta_0 = 2.548$.
 These are the values which are obtained with the help of statistical methods and are in
practice to use. For small $r$,
the deviations of $\alpha$ and $\beta$ with respect to their limiting values, $\alpha_0 = 
3.157$ and $\beta_0 = 2.548$, are very significant. With the increase of $r$, the deviations
decrease.

In the basis-set extrapolation, the limiting value $V_\infty$ ($n = \infty$) is expressed as
$V_\infty = A V_3 - (A -1) V_2$, where $A = 3^\gamma/(3^\gamma - 2^\gamma)$. Using the 
constant values $\alpha_0 = 3.157$ and $\beta_0 = 2.548$, we have calculated total energy
 $E_0$ at each position. For the values of $\alpha$ and $\beta$ at each position, we have 
calculated total energy $E$, and plotted in Figure \ref{Fig4}, the value of $(E - E_0)$ 
as a function of position. The values of $(E - E_0)$ are up to the order of $10^{-2}$
atomic unit (A U). The
 deviations shown in Figure \ref{Fig4} are quite significant, in particular for small $r$, as 
the total energy is calculated up to the accuracy of $10^{-8}$ A U.	

Finally, by using these values of $\alpha$ and $\beta$ at each position, the HF and EC, and 
finally the total energy, corresponding to $n = \infty$, can be calculated, and very 
accurate PES can be generated. 

\vspace{2mm}

\noindent
{\bf Acknowledgements}

\vspace{1mm}

Financial support from the Alexander von Humboldt Foundation, Germany, and the 
Department of Science \& Technology, New Delhi, India, and very helpful 
correspondence with Prof. Tatiana Korona are thankfully acknowledged. Suresh Chandra is
grateful to Prof. Dr. D.  Breitschwerdt and Prof. Dr. W.H. Kegel of Technical University,
 Berlin, Germany, for nice hospitality.

\vspace{3mm}

\noindent
{\bf References}
\begin{description}

\item{} 1. Truhlar, D.G., Chem. Phys. Lett. {\bf 294}, 45 (1998). 

\item{} 2. For example, Wheeler, M.D. \& Ellis, A.M. Chem. Phys. Lett.
 {\bf 374}, 392 (2003). 

\item{} 3. Sinclair, M.W., Fourikis, N., Ribes, J.C., Robinson, B.J., Brown, 
R.D. \& Godfrey, P.D., Austral. J. Phys. {\bf 26}, 85 (1973). 

\item{} 4. Werner, H.-J., Knizia, K.G., Manby, F.R., Sch\"utz, M., WIREs Comput Mol Sci 2, 
242-253 (2012), doi: 10.1002/wcms.82; MOLPRO, version 2012, a package of {\it ab initio} 
programs, Werner, H.-J., Knizia, K.G. {\it et al.}, see http://www.molpro.net.

\item{} 5. Pople, J.A., Head-Gordon, M. \& Raghavachari, K., J. Chem. Phys.
{\bf 87}, 5968 (1987).
\end{description}

\vspace{3mm}

\noindent
{\bf Author Contributions:} S.C. designed the research and wrote the paper. M.K.S. reviewed
 the literature and performed the calculations.

\vspace{3mm}

\noindent
{\bf Author Information:} Correspondence and requests for material should be addressed to 
S.C. (suresh492000@yahoo.co.in).
  


\begin{figure}[h]
\caption{\small Variation of $F$ [= $(4^{-\gamma} - 2^{-\gamma})/(3^{-\gamma} - 
2^{-\gamma})$] versus $\gamma$. The $F$ is plotted on the horizontal axis and $\gamma$ on the
 vertical axis.} \label{Fig1}
\end{figure}

\begin{figure}[h]
\caption{\small Values of $\alpha$ for various positions. We have $r$ = 2 (0.25) 6.5 \AA. 
Each $r$ has 28 positions.} \label{Fig2}
\end{figure}

\begin{figure}[h]
\caption{\small Values of $\beta$ for various positions. We have $r$ = 2 (0.25) 6.5 \AA. 
Each $r$ has 28 positions.} \label{Fig3}
\end{figure}

\begin{figure}[h]
\caption{\small Values of  $(E - E_0)$ for various positions. We have $r$ = 2 (0.25) 6.5 \AA.
Each $r$ has 28 positions.} \label{Fig4}
\end{figure}
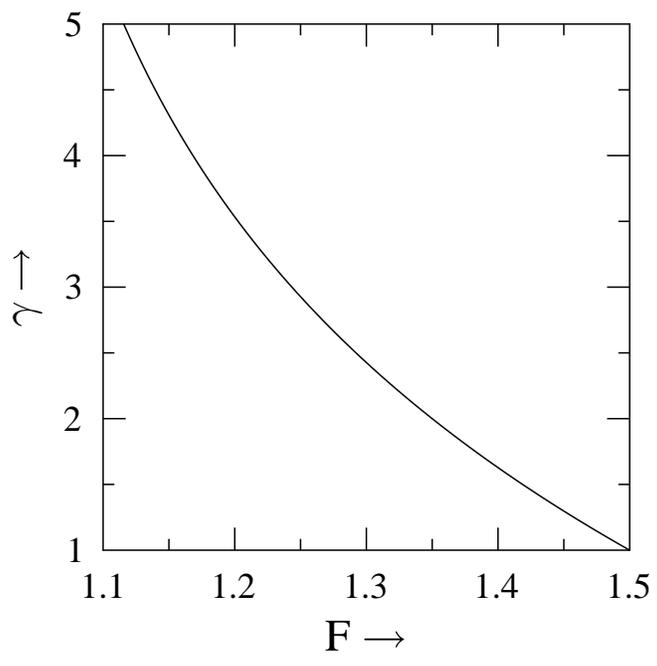
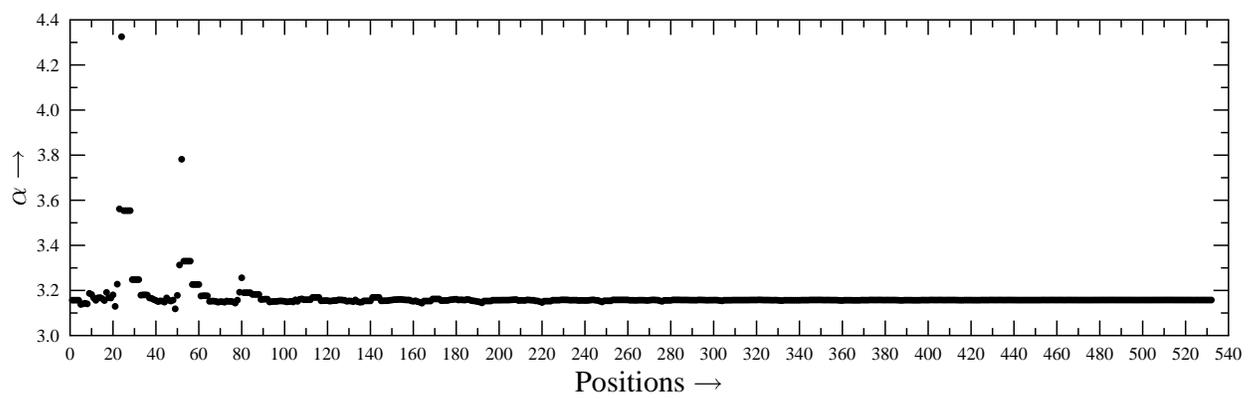
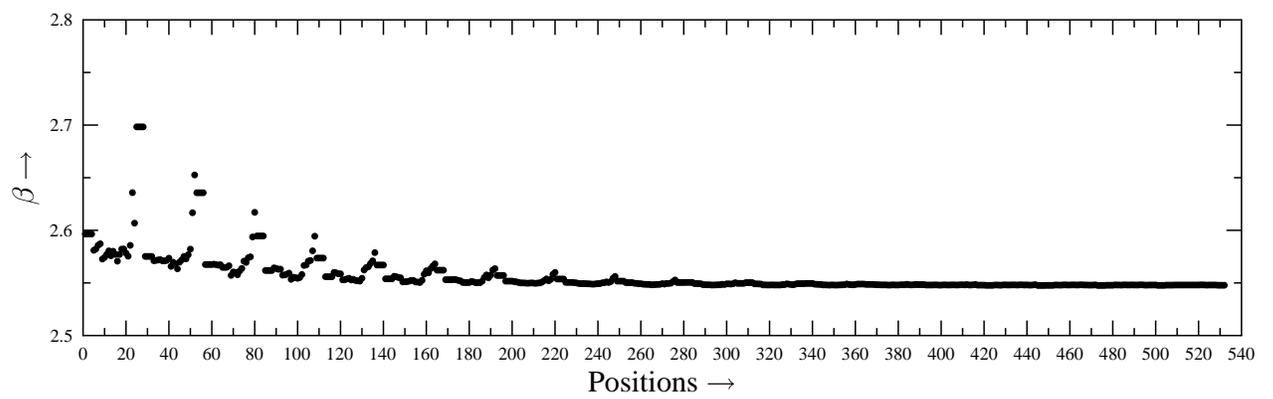
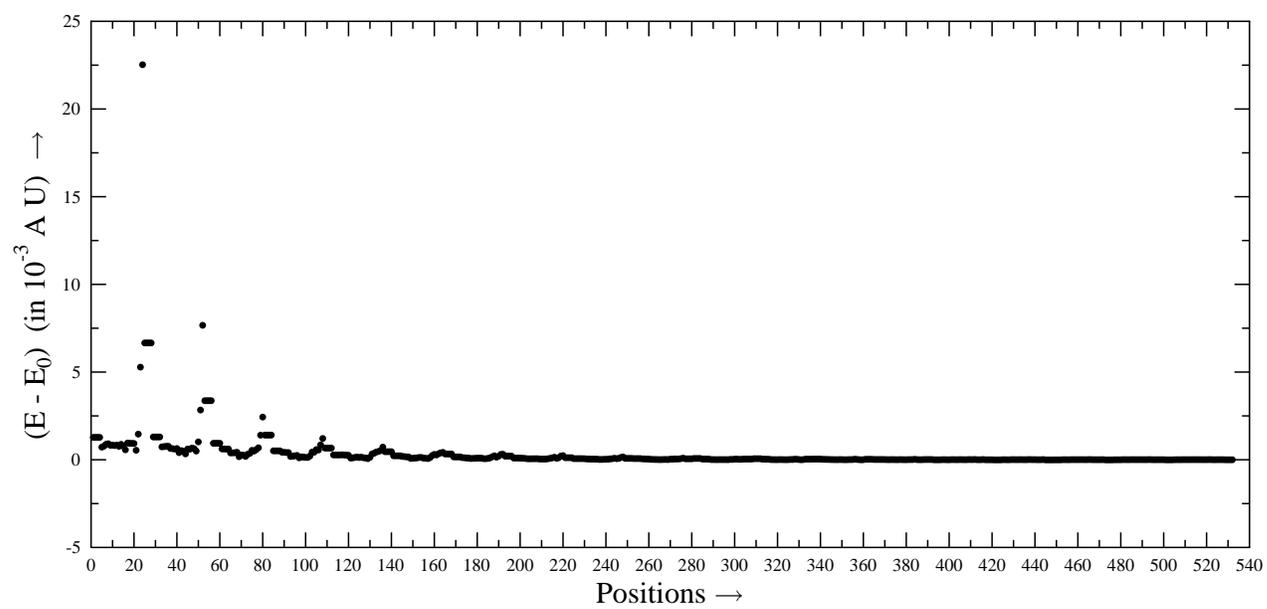
\end{document}